# Comprehensive segmentation of deep grey nuclei from structural MRI data

Manojkumar Saranathan[1], Giuseppina Cogliandro[2], Thomas Hicks[3], Dianne Patterson[3], Behroze Vachha[1], and Alberto Cacciola[2]

[1] Department of Radiology, UMass Chan Medical School, Worcester, MA

[2] University of Messina, Messina, Italy

[3] University of Arizona, Tucson, AZ

**Running head:** Deep grey nuclei segmentation
**Tables:** 2
**Figures:** 7

**\*Correspondence**

Manojkumar Saranathan, PhD
Department of Radiology
University of Massachusetts Chan Medical School
55 N Lake Avenue, Worcester, MA
Ph: (508)856-3558
Email: `manojkumar.saranathan@umassmed.edu`

# Abstract

Motivation: Lack of tools for comprehensive and complete segmentation of deep grey nuclei using a *single* software for reproducibility and repeatability

Goal(s): A fast accurate and robust method for segmentation of deep grey nuclei (thalamic nuclei, basal ganglia, claustrum, red nucleus) from structural T1 MRI data at conventional field strengths

Approach: We leverage the improved contrast of white-matter-nulled imaging by using the recently proposed Histogram-based Polynomial Synthesis (HIPS) to synthesize WMn-like images from standard T1 and then use a multi-atlas segmentation with joint label fusion to segment deep grey nuclei.

Results: The method worked robustly on all field strengths (1.5/3/7) and Dice coefficients of 0.7 or more were achieved for all structures compared against manual segmentation ground truth.

Impact: This method facilitates careful investigation of the role of deep grey nuclei by enabling the use of conventional T1 data from large public databases, which has not been possible, hitherto, due to lack of robust reproducible segmentation tools.

## Introduction

The role of subcortical brain structures is becoming increasingly important, both in healthy aging and, especially, in neurodegenerative and neuropsychiatric conditions. The involvement of the thalamus and its constituent nuclei have been shown in several conditions including multiple sclerosis [Minagar et al., 2013; Planche et al., 2020], frontotemporal dementia [Bocchetta et al., 2020; McKenna et al., 2022], Alzheimer's disease [Bernstein et al., 2021], and alcohol use disorder [Segobin and Pitel, 2021; Zahr et al., 2020]. Basal ganglia structures have been implicated in Parkinson's disease, Huntington's disease, dystonia and related conditions. Specific thalamic nuclei such as ventralis intermedius (VIM) and centromedian (CM) as well as basal ganglia structures such as the globus pallidus are powerful targets for deep brain stimulation (DBS) for treatment of drug-resistant essential and Parkinsonian tremors [7-8]. Newer subcortical DBS targets for treatment of intractable epilepsy [Klinger and Mittal, 2018; Ryvlin et al., 2021], obsessive compulsive disorder [Abraham et al., 2023; Zhang et al., 2020], and trigeminal neuralgia are topics of active research. Older indirect targeting methods are slowly being replaced by direct targeting and in this context, a method for fast, accurate, and comprehensive segmentation of subcortical structures is critical.

Many popular neuroimaging parcellation packages like Freesurfer [Fischl, 2012], FSL [Jenkinson et al., 2012], and volBrain [Manjón and Coupé, 2016]produce only a subset of subcortical structures (typically pallidum, nucleus accumbens, caudate, putamen, thalamus, hippocampus, amygdala for the left and right hemispheres) when segmenting 3D $T_1$-weighted MRI data. Freesurfer has also incorporated a thalamic nuclei segmentation module [Iglesias et al., 2018], which further subdivides the whole thalamus into constituent nuclei. Most of these packages lack subdivisions such as the internal and external divisions of the globus pallidus. They also do not segment structures such as the claustrum or the red nucleus. While there are specialized methods proposed to address parcellation some of the above structures, a seamlessly integrated unified package for segmentation of thalamic nuclei *and* subcortical structures is still missing. This would greatly improve repeatability and reproducibility efforts and minimize errors and overhead in installing and maintaining multiple packages across different operating systems.

Thalamus-optimized multi-atlas segmentation (THOMAS) [Su et al., 2019] is a current state-of-the-art technique for fast and accurate segmentation of thalamic nuclei from structural MRI. It leverages the improved contrast arising from white-matter nulling (WMn) [Saranathan et al., 2015; Tourdias et al., 2014] provided by sequences such as Fast Gray Matter Acquisition $T_1$ Inversion Recovery (FGATIR) and white-matter-nulled Magnetization Prepared Rapid Gradient echo (WMn-MPRAGE). WMn improves both intra-thalamic nuclear contrast as well as accurate delineation of the outer (ventral) boundaries of the thalamus from adjacent white matter tracts. THOMAS was originally developed and optimized for WMn contrast. However, it was very recently adapted for conventional $T_1$ (i.e. standard MPRAGE with dark cerebrospinal fluid signal) imaging by using a WMn synthesis step prior to segmentation [Vidal et al., 2024]. This preprocessing referred to as Histogram-based Polynomial synthesis (HIPS) essentially converts $T_1$ to WMn contrast, making it optimal for THOMAS, and greatly increasing the usability of THOMAS, especially on public databases such as Alzheimer's Disease Neuroimaging Initiative (ADNI) and Open access series of imaging studies (OASIS) which have only standard 3D T1 datasets.

In this work, we exploited the improved WMn contrast of THOMAS to segment other subcortical deep grey nuclei, including the claustrum and the red nucleus in addition to the thalamic nuclei. This method, we call sTHOMAS (s for subcortical), was validated on different image contrasts (WMn as well as standard $T_1$), different field strengths (3T and 7T MRI) as well as different manufacturers (GE, Siemens, Philips) to establish its accuracy and robustness. We expect the open-source containerized implementation to contribute both to repeatability and reproducibility efforts as well as systematic analysis of large databases to develop normative models and exploratory analyses.

## Methods

Multi-atlas generation: 20 7T WMn-MPRAGE datasets (the same 20 used in the original THOMAS implementation for thalamic nuclei) were segmented by an experienced neuroanatomist and a trained medical student using 3D-Slicer and ITK-SNAP. Nine subcortical structures- caudate, nucleus accumbens, claustrum, globus pallidus- interna and externa, whole globus pallidus, putamen, amygdala, and red nucleus- were manually delineated and added to the twelve structures (eleven thalamic nuclei and the mammillothalamic tract) that are already part of THOMAS for the left and right hemispheres. The list is shown in **Supplemental Table 1**.

Multi-atlas segmentation: The basic multi-atlas-based segmentation algorithm is described in detail in [Su et al., 2020] and shown in Figure 1. Briefly- the 20 WMn-MPRAGE prior datasets were diffeomorphically registered using Advanced Normalization Tools (ANTs) [Avants et al., 2008] to a mean WMn template (also generated from the 20 priors). These 20 transformations ($W_{piT}$) were precomputed and stored for computational efficiency. At runtime, the cropped mean WMn template is diffeomorphically registered to the cropped input image. Cropping, implemented using a simple automated scheme, was employed to improve accuracy and speed by restricting the focus to the structures of interest i.e. thalamus and deep grey nuclei. The crop size was chosen to encompass all the newer structures of interest which extend beyond the thalamus boundaries. By combining the two warps (i.e. prior to template $W_{piT}$ and template to input R), manually segmented labels from the 20 priors are warped into native image space and subsequently combined using a joint label fusion algorithm to yield the final parcellation labels. For standard $T_1$ data, the histogram-based polynomial synthesis (HIPS) algorithm was applied to the cropped image to synthesize WMn-like data prior to registration, and label fusion steps as described in [Vidal et al., 2024]. Note that the use of WMn synthesis enables the use of the more accurate joint label fusion algorithm compared to majority voting when combining the labels. The sTHOMAS pipeline is summarized in **Figure 1**. In addition to the nuclei for each hemisphere, sTHOMAS also outputs a quality control image file showing triplanar cross-sections of the input image with and without overlays of nuclei. It also has a panel where the registration between input and template is overlaid (edge map) for a quick visual evaluation of the efficacy of the registration step (**Supplemental Figure 1**). This file is useful for quick visual assessment when processing large databases. The volumes of each nucleus are also tabulated in separate tsv files, one for each hemisphere for use in statistical analyses. (**Supplemental Figure 2**)

Validation: sTHOMAS was rigorously validated for multiple use case scenarios. The first validation was a leave-one-out-cross-validation (LOOCV) scheme on the prior datasets using the 20 manual delineations of 7T WMn MPRAGE as gold standard. Following this, the accuracy of HIPS for processing T1 MRI was then validated on 18 T1 and WMn datasets obtained concurrently on subjects from Siemens and Philips 3T MRI scanners (different set of subjects scanned on each scanner). Finally, sTHOMAS was validated against manual segmentations on 25 3T standard T1 MRI datasets gleaned from the Human Connectome Project (HCP) young adult database. The manual segmentations were publicly available from [Rushmore et al., 2022]. For comparisons, FSL and Freesurfer were also used to segment subcortical structures in the same T1 datasets. Dice coefficients were used to assess accuracy. Note that since the accuracy of the

HIPS THOMAS segmentation for thalamic nuclei has already been previously validated and published [Vidal et al., 2024], this work only considers the newer subcortical structures added (with whole thalamus shown for reference purposes).

Results

**Figure 2** shows cropped axial sections from a 3T Philips WMn-MPRAGE dataset highlighting the improved intra-subcortical contrast with overlaid sTHOMAS segmentations. Note the clear delineation of the claustrum, globus pallidus interna and externa as well as small structures such as the mammillothalamic tract (MTT) and the habenula, not segmented by most currently available software. **Figure 3** shows sTHOMAS segmentation results from conventional 3T T1 MPRAGE datasets obtained on a Philips scanner (top row, same subject as shown in Figure 2) and a Siemens scanner (bottom row, different subject). Panels A and E show the original T1 contrast image whilst panels B and F show the corresponding HIPS synthesized WMn-like image. Note the clear visualization of small structures like in WMn-MPRAGE segmentation. **Figure 4** illustrates the robustness of the sTHOMAS method. The top row shows the results from a GE 1.5T conventional T1 dataset with slightly worse spatial resolution in axial and coronal sections. The bottom row shows the results from a 3T GE conventional T1 MPRAGE acquired on a patient with the semantic variant of frontotemporal dementia with clearly enlarged ventricles. Note that the nonlinear registration has performed well despite the challenges posed by the ventricles as can be seen by the accurate delineation of caudate nucleus and the thalamic structures adjacent to the ventricles such as mediodorsal and pulvinar nuclei.

Quantitative performance metrics for sTHOMAS are summarized in **Figure 5** for the LOOCV evaluated against manual segmentation ground truth for the newly added subcortical structures and the whole thalamus. Mean Dice values of 0.7 or higher was achieved for 8 of the 9 structures with three structures (caudate, putamen, thalamus) achieving Dice of 0.9 or higher. **Figure 6** illustrates the performance of sTHOMAS on conventional T1 MPRAGE evaluated against WMn-MPRAGE ground truth segmentation. The use of HIPS with sTHOMAS made it perform very comparably to that of WMn with a mean Dice of 0.8 or higher for all 9 structures. The corresponding data showing mean and standard deviation of the Dice values for both WMn and T1 MPRAGE are tabulated in Table 1. Finally, **Figure 7** compares the performance of sTHOMAS (green) with FSL (yellow) and Freesurfer (blue) on 3T T1 HCP datasets where manual segmentation was available for 5 structures (accumbens, caudate, pallidus, putamen, and thalamus). sTHOMAS was superior to Freesurfer for all except thalamus and superior or comparable to FSL for all except GP and thalamus. (see Discussion for explanation on the thalamus). The corresponding numerical Dice data with mean and standard deviation is shown in **Tables 1-2**.

## Discussion

We have developed sTHOMAS, a fast, comprehensive, robust, and accurate tool for subcortical segmentation that is open-source and publicly available in containerized form. Rigorous validation performed against manual segmentation ground-truth on 7TWMn and 3T T1 datasets indicates excellent accuracy (Dice) with almost all structures achieving Dice of 0.7 or more and structures like putamen, thalamus, and caudate achieving Dice of 0.8 and higher. We hope sTHOMAS will enable systematic analysis of large databases to throw more light on the role of subcortical structures like claustrum that have not been explored hitherto in healthy aging and disease. Compared to the previous releases of THOMAS, this is a lean implementation reducing the docker container size from 60Gb to16Gb making it much easier to download. It is also a fully open-source implementation with wrapper scripts provided for using docker and singularity containers, the latter being useful for high performance clusters where docker usage is restricted.

## Comparisons with existing methods

To our knowledge, apart from Freesurfer, there are no publicly available methods that can provide comprehensive deep grey nuclei segmentation which include individual thalamic nuclei (as opposed to whole thalamus). However, sTHOMAS possesses a significant computational time advantage over Freesurfer- 20-30 min vs several hours on an i9 processor 64 Gb RAM Linux workstation. THOMAS has also been shown to be slightly more accurate than Freesurfer and much more sensitive to atrophy in a recent work [Williams et al., 2024] rigorously comparing the two methods using Dice and Hausdorff distance as well as the ability to discriminate controls from mild cognitive impairment and Alzheimer's disease. In our comparisons on 3T HCP datasets against manuals segmentation, sTHOMAS also performed better than Freesurfer and better than or comparable to FSL. Other attempts at comprehensive deep grey nuclei segmentation include the Marseille 7T atlas [Brun et al., 2022] based on 7T MRI MP2RAGE datasets. This atlas is based on a single averaged template reducing its robustness. Finally, the Amsterdam group's Multi-contrast Anatomical Subcortical Structure Parcellation (MASSP) [Bazin et al., 2020] is a multi-atlas method which also incorporates shape in the models. While it produces 17 structures including STN and SN, the thalamus is not segmented to individual nuclei, which is a potential limitation. The striatum is also not separated into caudate and accumbens but instead segmented as a whole structure.

The claustrum is one of the most connected structures in the brain and one of the focus in Cricks' search [Crick and Koch, 2005] for structures involved in consciousness. It is still poorly understood, rarely segmented and its atrophy in various neurodegenerative conditions virtually undocumented. Recent evidence [Liaw and Augustine, 2023] suggests that the claustrum is potentially a structural interface between awakening, awareness, and integration, critical components of consciousness.

There are a couple of recently proposed specialized methods [Berman et al., 2020; Li et al., 2021] for segmentation of the claustrum including deep learning. Our accuracy is comparable or higher than these methods, making it a promising method to study the role of claustrum using a more widely distributed and used software tool.

Our method has several limitations. Due to the lack of widespread availability of susceptibility weighted MRI data in public databases, we have restricted the method to T1-weighted MRI (WMn or standard T1 MPRAGE contrast). As a result, two iron containing structures- subthalamic nucleus and substantia nigra- have been omitted. Future implementations will explore potential synthesis of susceptibility-weighted imaging (SWI) data from T1 and T2/T2* weighted MRI to segment these two missing nuclei. An immediate next step is the addition of ventricles and the whole hippocampus, which is currently in progress. Lastly, one potential drawback of using the Rushmore segmentation of HCP 3T T1 data is *their* lack of WMn MPRAGE for delineation of thalamus boundaries. This could explain the reason why sTHOMAS paradoxically performed slightly worse than FSL or Freesurfer for the whole thalamus. Note that 7T WMn-MPRAGE data which leveraged WMn contrast for delineation of the thalamus boundaries, produced much higher sTHOMAS Dice for the whole thalamus (0.92 for WMn vs manual, 0.93 for T1HIPS vs WMn, and 0.81 for T1 vs manual Rushmore labels). Nonetheless, the Rushmore dataset employs state-of-the-art unified procedures for comprehensive segmentation and was, hence, included to evaluate sTHOMAS against FSL and Freesurfer on standard T1 MRI data.

Future enhancements besides increasing the number of structures include incorporation of deep learning-based segmentation which can further improve computational speed and reduce errors from misregistration resulting from enlarged ventricles and potentially compromising the accuracy of caudate segmentation.

In conclusion, sTHOMAS can provide accurate and fast segmentation of deep grey structures from T1 MRI data, working robustly across different field strengths and image contrasts. We expect sTHOMAS to be useful in fast analysis of public databases to uncover the roles of these structures in healthy aging and disease.

## Acknowledgements

We would like to acknowledge funding from the National Institute of Biomedical Imaging and Bioengineering (R01 EB032674). Data were provided [in part] by the Human Connectome Project, WU-Minn Consortium (Principal Investigators: David Van Essen and Kamil Ugurbil; 1U54MH091657) funded by the 16 NIH Institutes and Centers that support the NIH Blueprint for Neuroscience Research; and by the McDonnell Center for Systems Neuroscience at Washington University. We also acknowledge use of a 1.5T MRI dataset from the publicly available Indian Brain Segmentation Database [Jayanthi Sivaswamy et al., 2021].

Figures

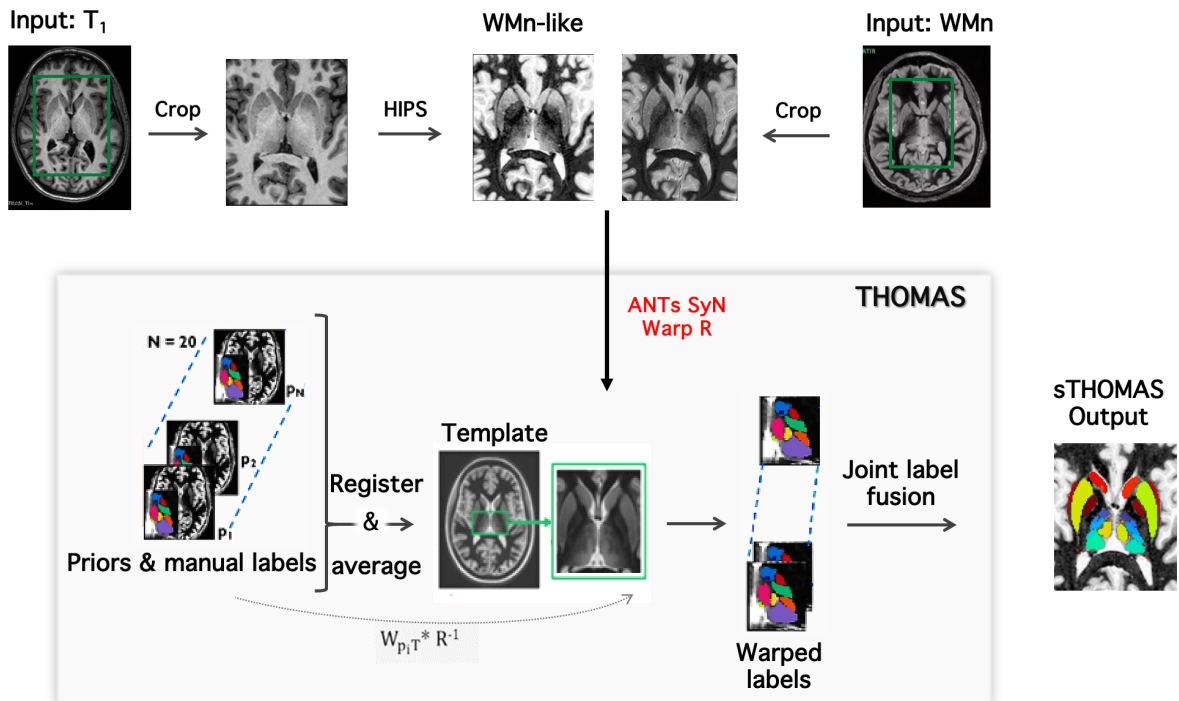

Figure 1. The sTHOMAS segmentation pipeline. The input image is cropped and undergoes an additional HIPS step for synthesis of WMn-like images in the case of standard T1 and then registered nonlinearly to a cropped internal template image. This transformation is combined with the precomputed prior-template warps to warp the prior labels to native space which are then fused using a joint-label fusion algorithm to generate the final sTHOMAS segmentation outputs.

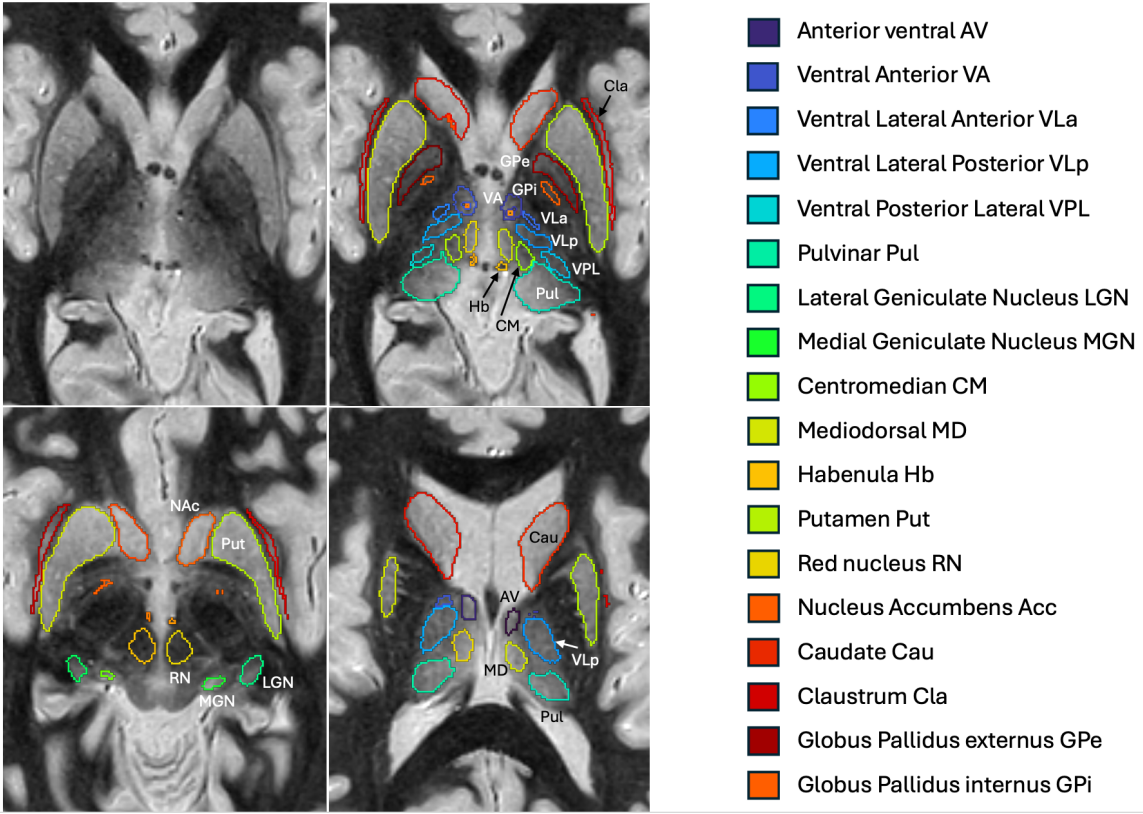

Figure 2. Axial sections from a Philips 3T WMn-MPRAGE dataset with bilateral sTHOMAS segmentation outputs overlaid. Note the clear visualization of small structures such as the lateral and medial geniculate nuclei, habenula and the mammillothalamic tract as well as structures like the claustrum and red nucleus not segmented by most existing segmentation tools.

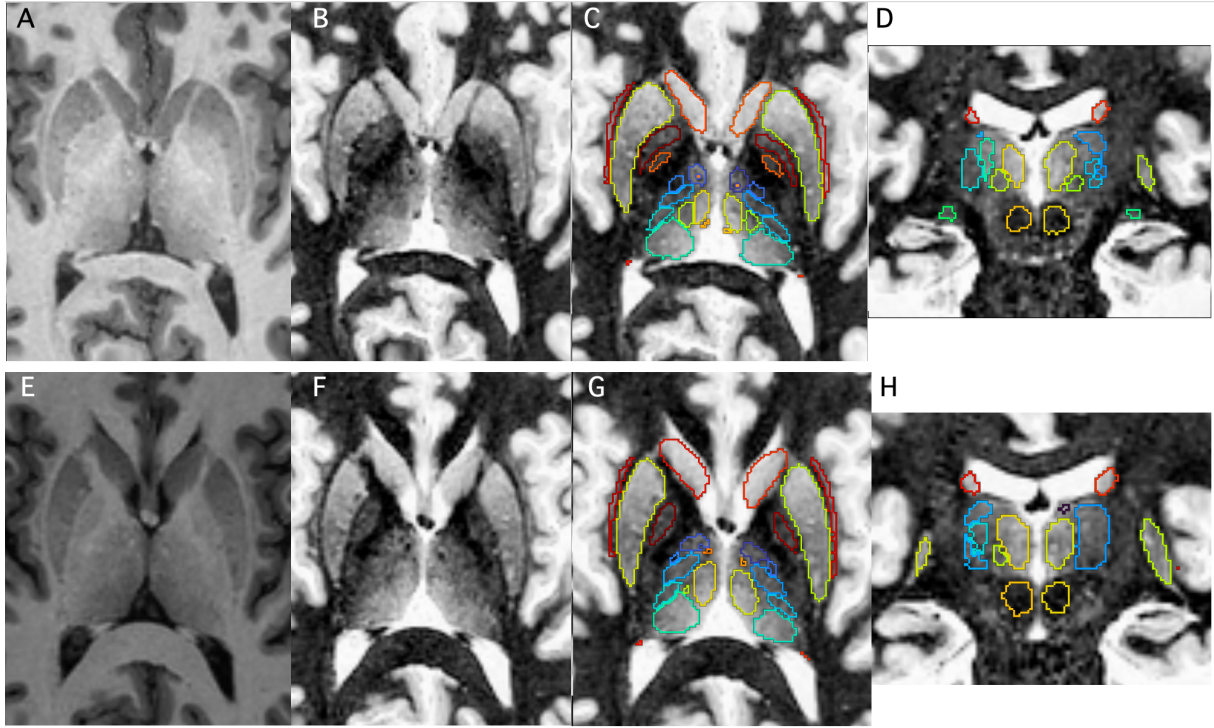

Figure 3. sTHOMAS segmentation results from a Philips 3T T1-MPRAGE dataset (top row) and a Siemens 3T T1 MPRAGE dataset (bottom row) showing original T1 input images (A,E) and the corresponding HIPS-synthesized WMn-like images (B,F) and with the sTHOMAS overlays (C,G). Coronal cuts with overlays are shown in the rightmost column (D,H).

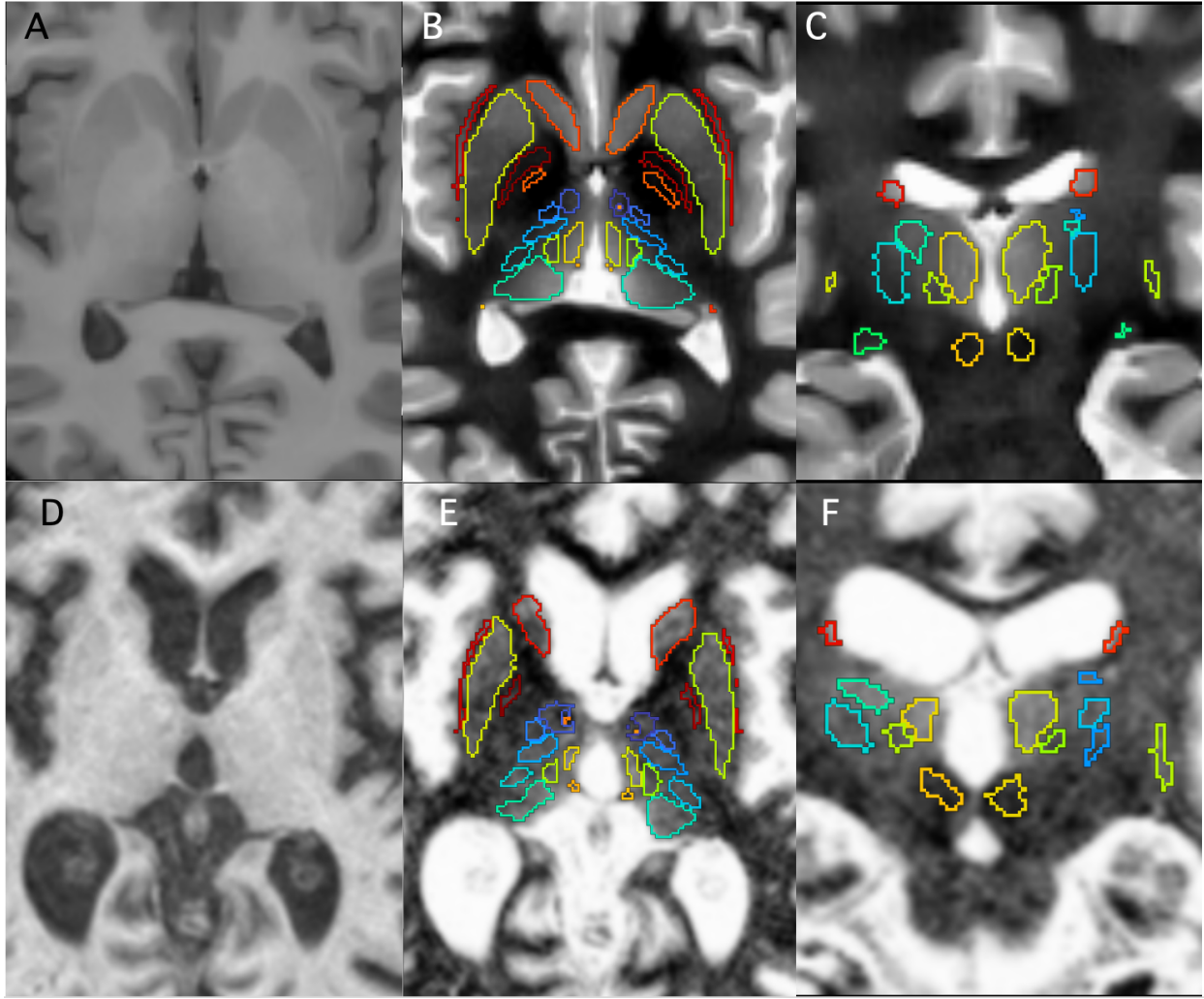

Figure 4. sTHOMAS segmentation results from a GE 1.5T T1-MPRAGE dataset (top row) and a GE 3T T1 MPRAGE dataset on a patient with the semantic variant of frontotemporal demential (bottom row) showing original T1 input images (A,B) and the corresponding HIPS-synthesized WMn-like images with sTHOMAS overlays (B,E). Coronal cuts with overlays are shown in the rightmost column (C,F). Note the excellent performance of sTHOMAS even on the patient with enlarged ventricles and minimal motion artifacts.

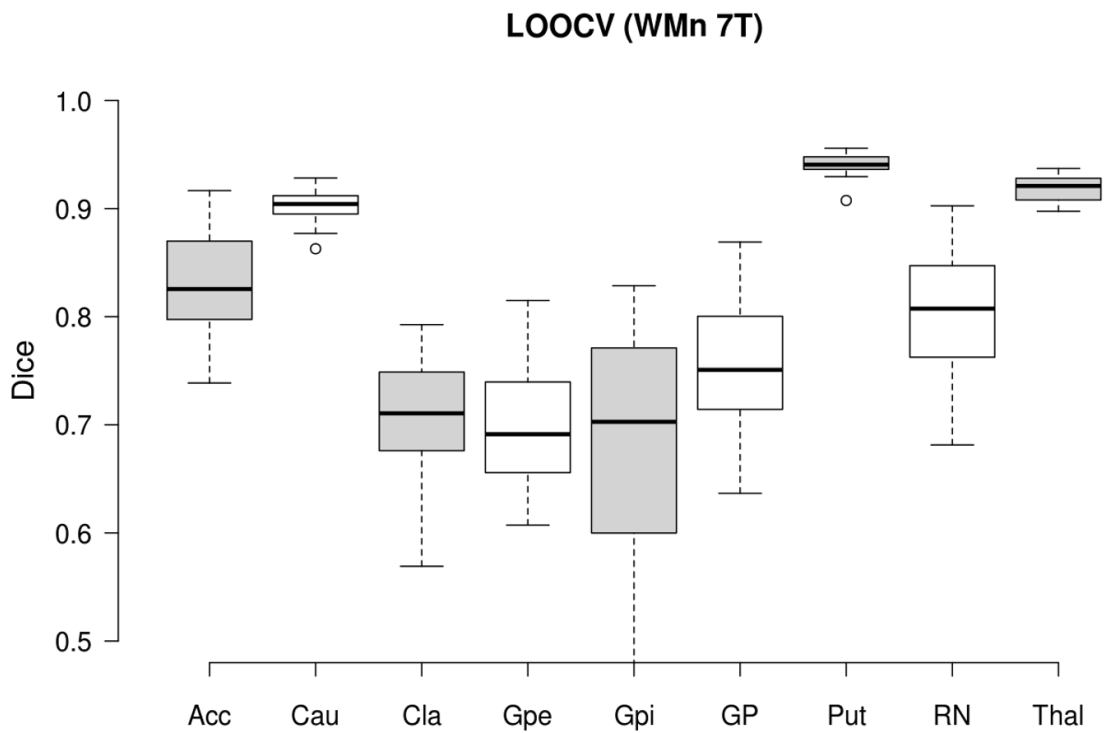

Figure 5. Mean Dice results (n=20) from GE 7T WMn-MPRAGE data compared against manual segmentation ground truth using a leave-one-out-cross-validation (LOOCV) approach.

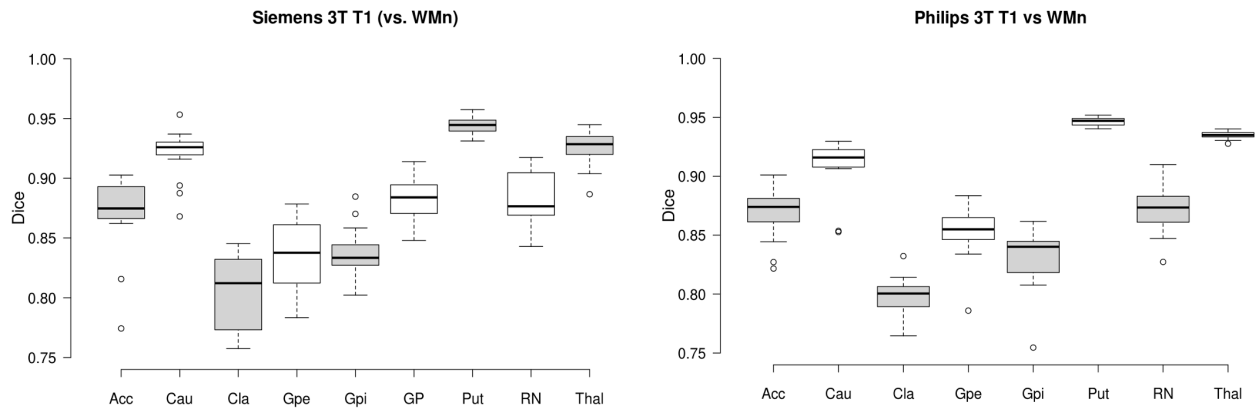

Figure 6. Mean Dice results (n=18) from Siemens 3T T1-MPRAGE data (left) and Philips 3T T1-MPRAGE data (right) compared against the corresponding WMn-MPRAGE data acquired on the same subject. The high Dice values show the validity of the HIPS synthesis step used in sTHOMAS for processing standard T1 MRI data.

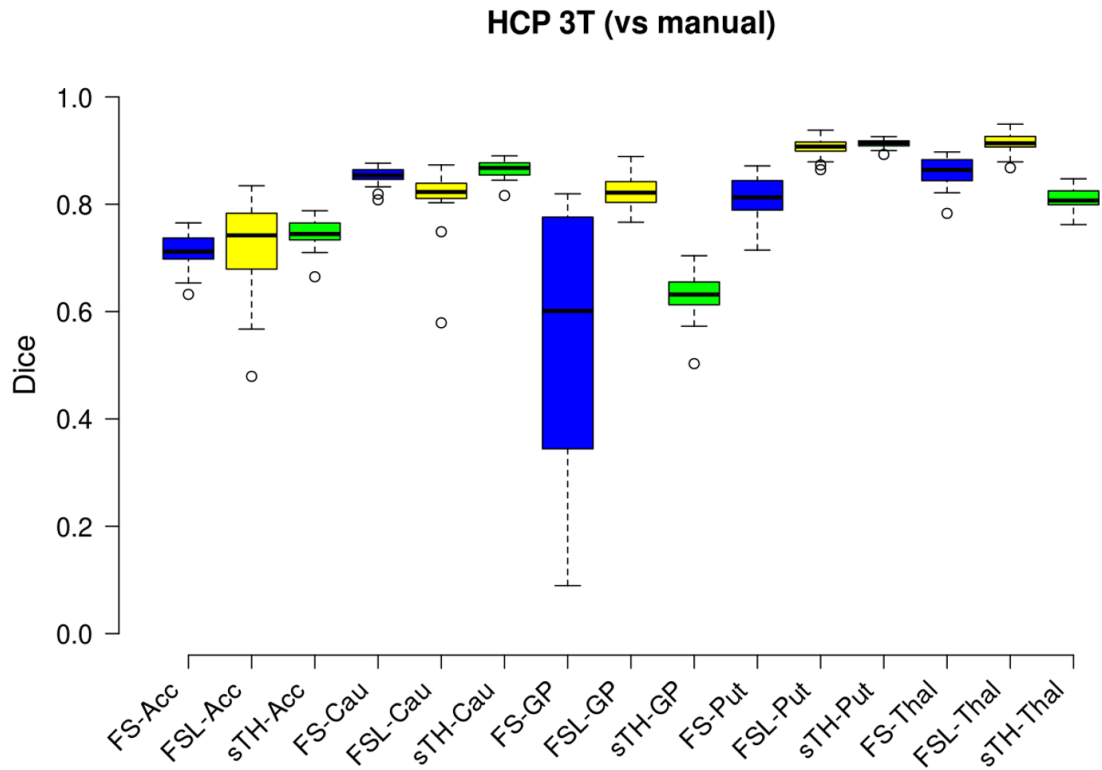

Figure 7. Mean Dice results (n=25) from the HCP 3T T1-MPRAGE data for sTHOMAS (green) compared to Freesurfer (blue) and FSL (yellow). Only structures common to sTHOMAS and the Rushmore manual segmentation labels were used for the evaluation.

Table 1. Mean and standard deviation of Dice results for sTHOMAS evaluated on GE 7T WMn MPRAGE (n=20) against manual segmentation. The right columns show the sTHOMAS Dice for Philips and Siemens 3T T1 MPRAGE (n=18) evaluated against WMn-MPRAGE i.e. evaluation of HIPS accuracy.

| Structure | 7T WMn GE | 3T $T_1$ Philips | 3T $T_1$ Siemens |
|---|---|---|---|
| Accumbens | 0.83 +/- 0.05 | 0.87 +/- 0.02 | 0.87 +/- 0.03 |
| Caudate | 0.90 +/- 0.02 | 0.91 +/- 0.02 | 0.92 +/- 0.02 |
| Claustrum | 0.71 +/- 0.06 | 0.80 +/- 0.02 | 0.79 +/- 0.07 |
| Globus pallidus ext | 0.70 +/- 0.06 | 0.85 +/- 0.02 | 0.84 +/- 0.03 |
| Globus pallidus int | 0.67 +/- 0.14 | 0.83 +/- 0.02 | 0.84 +/- 0.02 |
| Globus pallidus | 0.76 +/- 0.06 |  | 0.88 +/- 0.02 |
| Putamen | 0.94 +/- 0.01 | 0.95 +/- 0.00 | 0.94 +/- 0.01 |
| Red nucleus | 0.80 +/- 0.06 | 0.87 +/- 0.02 | 0.88 +/- 0.02 |
| Thalamus | 0.92 +/- 0.01 | 0.93 +/- 0.00 | 0.93 +/- 0.01 |

Table 2. Mean and standard deviation of Dice results on HCP 3T T1 dataset (n=25) evaluated against manual segmentation comparing Freesurfer and FSL with sTHOMAS.

| Structure | Freesurfer | FSL | sTHOMAS |
| --- | --- | --- | --- |
| Accumbens | 0.71 +/- 0.03 | 0.72 +/- 0.08 | **0.74 +/- 0.03** |
| Caudate | 0.85 +/- 0.02 | 0.82 +/- 0.06 | **0.87 +/- 0.02** |
| Globus pallidus | 0.54 +/- 0.25 | **0.82 +/- 0.03** | 0.63 +/- 0.04 |
| Putamen | 0.81 +/- 0.04 | **0.91 +/- 0.02** | **0.91 +/- 0.01** |
| Thalamus | 0.86 +/- 0.03 | **0.91 +/- 0.02** | 0.81 +/- 0.02 |

Supplemental material

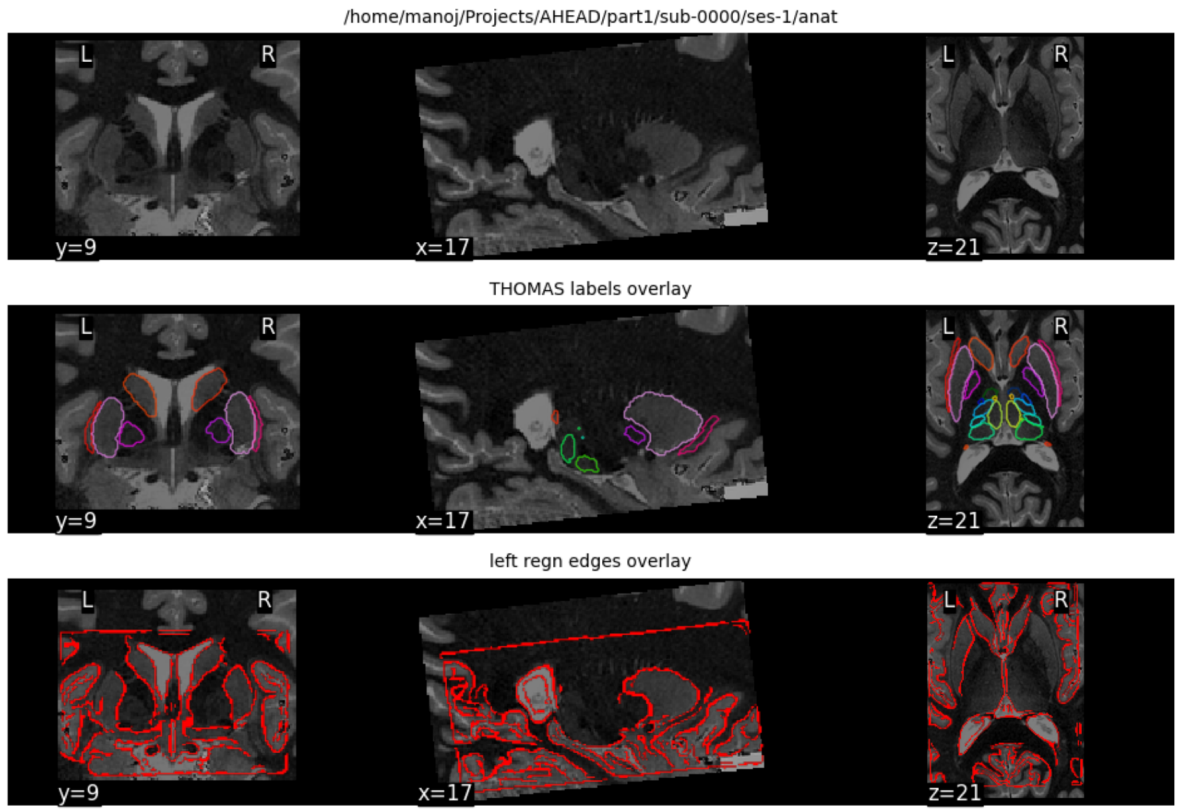

Supplemental Figure 1. An example of an automatically generated output file (png format) from sTHOMAS pipeline for quality control purposes showing the original cropped image in all 3 orthogonal planes (top), with sTHOMAS bilateral outputs overlaid (middle), and the edge map of registration with the WMn template overlaid (bottom). The edge map is useful to rule out registration failures occasionally seen in cases with enlarged ventricles especially near the caudate nucleus.

```
1-THALAMUS 6431.000000
2-AV 139.000000
4-VA 331.000000
5-VLa 94.000000
6-VLP 973.000000
7-VPL 360.000000
8-Pul 1544.000000
9-LGN 143.000000
10-MGN 81.000000
11-CM 122.000000
12-MD-Pf 683.000000
13-Hb 26.000000
14-MTT 38.000000
26-Acc 761.000000
27-Cau 3657.000000
28-Cla 924.000000
29-GPe 392.000000
30-GPi 219.000000
31-Put 5736.000000
32-RN 232.000000
33-GP 717.000000
34-Amy 213.000000
```

Supplemental Figure 2. Content of an exemplary nucleiVols.txt generated for each hemisphere listing the volumes of whole thalamus and the segmented nuclei in mm$^3$

Supplemental Table 1. List of structures segmented by sTHOMAS classified into regions (thalamus, basal ganglia, other). The abbreviation/label index followed for the output nomenclature is also shown.

| Region | Structure/nucleus | Abbreviation label index |
| --- | --- | --- |
| Thalamus (Anterior) | Anteroventral | AV 2 |
| Thalamus (Ventral) | Ventral anterior | VA 4 |
| | Ventral lateral anterior | VLa 5 |
| | Ventral lateral posterior | VLp 6 |
| | Ventral posterolateral | VPL 7 |
| Thalamus (Posterior) | Pulvinar | Pul 8 |
| | Lateral geniculate nucleus | LGN 9 |
| | Medial geniculate nucleus | MGN 10 |
| Thalamus (Medial) | Centromedian | CM 11 |
| | Mediodorsal-parafascicular | MD-Pf 12 |
| Basal Ganglia | Caudate | Cau 26 |
| | Nucleus accumbens | Acc 27 |
| | External globus pallidus | GPe 29 |
| | Internal globus pallidus | GPi 30 |
| | Putamen | Put 31 |
| | Globus pallidus whole | GP 33 |
| Other | Habenula | Hb 13 |

|  | Mammillothalamic tract | MTT 14 |
|  | Claustrum | Cla 28 |
|  | Red nucleus | RN 32 |